\documentstyle[bbm,amsfonts,epsfig,12pt,here]{article}

\newcommand {\eqref} [1] {(\ref {#1})}
\newcommand {\slsh} [1] {\not{\hbox{\kern-2pt${#1}$}}}

\def\drawbox#1#2{\hrule height#2pt
         \hbox{\vrule width#2pt height#1pt \kern#1pt
               \vrule width#2pt}
               \hrule height#2pt}

\def\Asym#1#2{\vcenter{\vbox{\drawbox{#1}{#2}
               \kern-#2pt       
               \drawbox{#1}{#2}}}}

\def\asymm{\Asym{6.4}{0.3}}

\def\basymm{\overline{\asymm}}

\newcommand {\beq} {\begin{equation}}
\newcommand {\eeq} {\end{equation}}
  \newcommand {\ber}{\begin{eqnarray*}}
  \newcommand {\eer} {\end{eqnarray*}}
\newcommand {\bea}{\begin{eqnarray}}
  \newcommand {\eea} {\end{eqnarray}}

\newcommand{\None}{${\cal N}=1\ $}

\def\Acknowledgements{\bigskip  \bigskip {\begin{center} \begin{large}
              \bf ACKNOWLEDGMENTS \end{large}\end{center}}}

\begin{document}
\begin{titlepage}
\begin{flushright}{CERN-TH/2003-201

FTPI-MINN-03/23, UMN-TH-2212/03}

\end{flushright}
\vskip 1cm

\centerline{{\Large \bf QCD Quark Condensate from }}
\vskip 0.1cm
\centerline{{\Large \bf SUSY and the Orientifold Large-$N$ Expansion}}
\vskip 1cm
\centerline{\large A. Armoni ${}^a$, M. Shifman ${}^{b}$, G. 
Veneziano ${}^{a,c}$}
\vskip 0.1cm

\vskip 0.5cm
\centerline{${}^a$ Theory Division, CERN}
\centerline{CH-1211 Geneva 23, Switzerland}
\vskip 0.5cm
\centerline{${}^b$ William I. Fine Theoretical Physics Institute, 
University
of Minnesota,}
\centerline{Minneapolis, MN 55455, USA}
\vskip 0.5cm
\centerline{${}^c$ Kavli Institute for Theoretical Physics,}
\centerline{University of California, Santa Barbara, CA 93106, USA}
\vskip 1cm

\begin{abstract}

We estimate  the quark condensate in one-flavor massless QCD from the known value of the gluino
condensate   in SUSY Yang-Mills theory
using our newly proposed ``orientifold"
large-$N$ expansion. 
 The numerical result for the quark condensate renormalized at the scale  $2~\rm{GeV}$ is then given as a function of $\alpha_s(2\,\, \rm{GeV})$ and of possible  corrections from sub-leading terms.  Our value can be compared with the quark condensate in 
(quenched) lattice  QCD or with the one extracted from the
Gell-Mann--Oakes--Renner  relation by virtue of non-lattice
determinations of the quark masses.
 In both cases we find quite a remarkable agreement. 

\end{abstract}

\end{titlepage}

\section{Introduction}
\label{introduction}

\noindent

In non-supersymmetric theories, such as QCD, it is very difficult,
if possible at all,  to perform reliable analytic
calculations in the non-perturbative regime. 
In the supersymmetric version of the theory the
situation is much better, due to holomorphy. In particular the {\it
exact} gluino condensate \cite{gluinocond}
can be evaluated in \None Super-Yang-Mills.

In previous publications \cite{ASV1,ASV2} we suggested a precise way of copying
non-perturbative results from a supersymmetric theory to a certain
non-supersymmetric theory named {\em orientifold field
theory} (due to its realization via orientifold type-0 string
theory \cite{Sagnotti:1995ga}). The orientifold theory is an SU$(N)$ gauge 
theory coupled
to a Dirac fermion in the two-index antisymmetric representation,
$\asymm + \basymm$. Similarly, one can consider \cite{Armoni:2003jk}
a generalized orientifold
QCD (or ${\rm QCD}_{\rm OR}$), which consists of $N_f$ flavors of Dirac
fermions in the antisymmetric representation $(N_f\geq 1)$.

Our purpose here is to carry out an explicit  calculation
 of the quark condensate in
one-flavor QCD  anticipated in Ref.~\cite{ASV2}.
Let us briefly recall the idea behind such a calculation. Consider 
three one-parameter families 
of gauge theories, the above-mentioned parameter being  $N$, of their common gauge group SU$(N)$:

\begin{itemize}
\item Pure Yang Mills (YM) theory also known as gluodynamics;
\item   
$\rm{QCD}_{\rm{F}}$ i.e.  
 standard  't Hooft's  extension of QCD at arbitrary $N$  (the number of quarks in the fundamental plus anti-fundamental  representation $N_f$
 is kept fixed);
\item  $\rm{QCD}_{\rm{A}}$, i.e. the SU$(N)$ gauge theory  with $N_f$ {\em Majorana} fermions in the {\em adjoint representation}.
\end{itemize}

In Ref.~\cite{ASV2} we  made a simple observation that, as $N$  
increases,  the generalized orientifold theory $\rm{QCD}_{\rm{OR}}$ 
interpolates between the  three other theories above. Indeed,
at $N=2$,  the fermions of $\rm{QCD}_{\rm{OR}}$ are gauge singlets, and, 
therefore,   $\rm{QCD}_{\rm{OR}}$ reduces to YM. At $N=3$, $\rm{QCD}_{\rm{OR}}$ obviously coincides with  $\rm{QCD}_{\rm{F}}$. 
Moreover, at $N \rightarrow \infty$
the bosonic sector  of  $\rm{QCD}_{\rm{OR}}$ goes into that 
of $\rm{QCD}_{\rm{A}}$ --- a straightforward generalization 
\cite{Armoni:2003jk} of the results of  Ref.~\cite{ASV1}.
This last observation becomes particularly interesting for one massless 
flavor, since in this case the limiting large-$N$ theory 
$\rm{QCD}_{\rm{A}}$ is nothing but the supersymmetric generalization 
of YM theory --- SYM theory also known
as supersymmetric gluodynamics. We will limit our attention to 
this case hereafter.

A consistency check of the above statements can be made by comparing the 
coefficients of the $\beta$ functions of these theories at different values 
of $N$, as well as the anomalous dimensions of the corresponding fermion bilinears $\bar{\psi} \psi$.  In Table 1 we present this check for the  $N_f=1$ case 
under discussion.

\vspace{5mm}

\begin{center}
\begin{tabular}{|c|c|c|c|c|}
\hline
$\frac{\mbox{ Theory} \to  }{\mbox{  Coefficients} \downarrow} $ \rule{0cm}{0.7cm} &YM&$\rm{QCD}_{\rm{F}}$& $\rm{QCD}_{\rm{OR}}$& SYM \\[3mm]
\hline\hline
\rule{0cm}{0.7cm}$\beta_0$  & $\frac{11}{3}\, N$ & $\frac{11}{3}\, N-\frac{2}{3}$&$3N+\frac{4}{3}$ &$3N$\\[2mm]
\hline   
\rule{0cm}{0.7cm}$\beta_1$ &$\frac{17}{3}\, N^2$ & $\frac{17}{3}N^2-\frac{13}{6}N+\frac{1}{2N}$&
$3N^2 +\frac{19}{3}N-\frac{4}{N}$ &$3N^2$\\[2mm]
\hline
\rule{0cm}{0.7cm}$\gamma$ & $\star$   & $\frac{3(N^2-1)}{2N}$&$\frac{3(N-2)(N+1)}{N}$ &$3N$\\[2mm]
\hline

\end{tabular}

\vspace{4mm}

Table 1. 

\end{center}

\noindent
We use the standard definition of the coefficients
of the $\beta$
function from PDG, see Ref. \cite{Hinchliffe},
\beq
\mu\,\frac{\partial\alpha}{\partial \mu}\equiv 2\beta (\alpha)
=-\frac{\beta_0}{2\pi } \alpha^2 -\frac{\beta_1}{4\pi^2 } \alpha^3 +...
\label{gmlf}
\eeq
The coefficients can be found in \cite{Tarasov:1980kx}, where formulae 
up to three loops are given. The  anomalous dimension $\gamma$
of the fermion bilinear operators $\bar\Psi\Psi$ is normalized in such a way that
\beq
\left(\bar\Psi\Psi\right)_Q = \kappa^{\gamma/\beta_0}\,\left(\bar\Psi\Psi\right)_\mu\,,~~\kappa \equiv \frac{\alpha (\mu)}{\alpha (Q)}\,,
\eeq
and $\mu$ and $Q$ denote the normalization points. For our present 
purposes  we can limit ourselves to the two-loop
$\beta$-functions and the one-loop anomalous dimensions. We can check
easily that the various coefficients of
$\rm{QCD}_{\rm{OR}}$ go smoothly from those of YM ($N=2$) through
those of $\rm{QCD}_{\rm{F}}$ ($N=3$) to those of SYM theory 
at $N \rightarrow \infty$.

Since many non-perturbative properties of SYM 
theory are known, the large-$N$ expansion will provide us with information 
on the non-perturbative behavior of one-flavor 
QCD at $N=3$,  modulo $1/N \sim 1/3$ corrections.
The planar equivalence method is applicable
to  a large class of bosonic correlators and can be tested,
in principle, in lattice calculations. In this letter we will concentrate 
our attention on a one-point function, the quark condensate,
which has been explicitly computed in SYM theory, and 
measured on the lattice.

\section{Renormalization-group-invariant quantities with a smooth 
large-$N$ limit}
\label{renormalization}

\noindent

In order to carry out our calculation we need to define, for each theory, 
renormalization-group-invariant (RGI) fermion-bilinear operators and 
compare their vacuum expectation values 
(VEV's) with the appropriate power of the corresponding fundamental 
RGI scale $\Lambda$. Furthermore, we would like always to deal  
with quantities that can be expanded at large $N$ and fixed 
't~Hooft ($N$-independent) coupling $\lambda$,
\beq
\lambda \equiv \frac{g^2 N}{8 \pi^2} = \frac{\alpha N}{ 2 \pi}\,,
\label{thooftc}
\eeq
 as a (possibly asymptotic) power series in $1/N$. It turns out that this latter requirement calls for a slightly unconventional definition of the above quantities since at fixed $\lambda$
 a non-integer power of $g^2$ behaves  as a non-integer power of $N$ and is, 
 thus,  non-analytic at $N = \infty$.

The 
standard (two-loop) definition of the scale parameter $\Lambda$,
which follows from the conventions of Ref.~\cite{Hinchliffe},
is
\beq
\Lambda_{\rm{standard}} =\mu \left(\frac{16\pi^2}{\beta_0 \, g^2(\mu)} 
\right)^{\beta_1/\beta_0^2}
\exp\left( - \frac{8\pi^2}{\beta_0 \, g^2 (\mu)}  
\right) \,.
\label{lam}
\eeq
(For further details see Appendix A.)
Because  $g^2$ is multiplied everywhere by $\beta_0$, Eq.~(\ref{lam}) does 
not suffer from the above-mentioned problem; it defines an 
$N$-independent constant of dimension of mass.
In what follows, it will be convenient to adopt a
more general definition,
\beq
\Lambda_{c}=\mu \left( c \,\lambda (\mu) \right)^{-\beta_1/\beta_0^2}
\,\,
\exp\left( - \frac{N}{\beta_0}\,\,\frac{1}{\lambda (\mu )} \,  
\right)\, ,
\label{lamus}
\eeq
where the constant $c$ has a finite large-$N$ limit around which it can be expanded.  In the standard definition $c=\beta_0/(2N)$, cf. Eq.~(\ref{lam}).
For the time being we will keep $c$ as a free parameter, and  
will discuss  the sensitivity of our results to the choice of $c$ later. 
In a similar manner we introduce RGI bifermion operators  as\,\footnote{
The kinetic term of the fermion fields is canonically normalized,
i.e. ${\cal L} = \bar\Psi\not \!\! D\Psi $.}
\beq
 \langle \bar\Psi \Psi \rangle_{\tilde{c}}  \equiv N^{-2} \, \left( \tilde{c} \,\lambda (\mu ) \right)^{\gamma/\beta_0} \,  \langle \bar\Psi \Psi  \rangle\,,
\label{RGicond}
\eeq
where $\tilde{c}$, like $c$, has a smooth  large-$N$ limit.  Its impact on
$ \langle \bar\Psi \Psi \rangle_{\tilde{c}} $ will be 
 discussed later. Equation (\ref{RGicond})
 will be applied both to the gluino condensate in SYM theory and to the
 quark condensate in the orientifold theory.
With the above definitions  the condensates  and $\Lambda$'s  approach finite limits as
$N \rightarrow \infty$. Moreover, the ratio
\beq
R(N) =  \frac{\langle \bar\Psi \Psi \rangle_{\tilde c}}{ \Lambda_{c}^3}
\label{ratios}
\eeq
 also approaches a finite limit at large $N$ in both theories,
 and enjoys a smooth $1/N$ expansion.

\section{The gluino condensate and the large-$N$ limit of the orientifold condensate}
\label{gluinoc}

\noindent

In SYM theory, where a number of exact results were obtained,
the general considerations of Sect.~\ref{renormalization}  simplify considerably. First of all, the expression for
$\Lambda$ in Eq.~(\ref{lamus}) becomes {\em exact} \cite{nsvzbeta}
rather than the two-loop approximation, 
\beq
\Lambda^3_{\rm SYM} =\mu^3 
\left(\frac{1}{c \, \lambda (\mu )}\right) \,\exp\left(- \frac{1}{\lambda (\mu )}\right) \,.
\eeq
(In this case the standard value of $c =3/2$.)
Furthermore, the definition (\ref{RGicond}) of the gluino condensate becomes
\beq
 \langle \lambda \lambda \rangle_{\tilde{c}} \equiv \frac{1}{N^2} \, \left( \tilde{c} \,\lambda (\mu )\right)  \,\left\langle 
\lambda^{a\,,\alpha} \lambda^a_{\alpha}  \right \rangle \,.
\label{RGglcond}
\eeq
Note that we deal here with the holomorphic part, there is no complex conjugate term in the right-hand side. The reason is that the exact results for the gluino condensate are routinely presented  in terms of  the holomorphic condensate, see below. Generally speaking, the VEV $\left\langle 
\lambda^{a\,,\alpha} \lambda^a_{\alpha}  \right \rangle $
is complex. We will assume the vacuum angle $\theta$ to vanish.
Then one can choose the vacuum state in such a way that
$\left\langle 
\lambda^{a\,,\alpha} \lambda^a_{\alpha}  \right \rangle $
is real. We will discuss shortly how the gluino condensate
defined in this particular way is mapped onto the orientifold
theory.

The exact expression for the gluino condensate
in SU($N$) supersymmetric gluodynamics can be obtained
from  weak coupling considerations  \cite{Novikov:ic}.
All numerical factors are carefully collected for SU(2) in the review paper
\cite{Shifman:1999mv}. A weak coupling calculation for SU($N$)
with arbitrary $N$ was carried out in \cite{Davies:1999uw}.
Note, however, that an unconventional definition of the scale parameter 
$\Lambda$ is used in Ref.~\cite{Davies:1999uw}. One can pass to
the conventional definition of $\Lambda$ either by normalizing the result
to the SU(2) case \cite{Shifman:1999mv} or by analyzing the context 
of Ref.~\cite{Davies:1999uw}. Both methods give the same result,
see Appendix A. When   expressed in terms of our
$\Lambda_{\rm SYM}$ it gives, for the ratio defined in Eq.~(\ref{ratios}) 
\beq
R_{\rm SYM}(N) = - \frac {c \tilde{c}}{2\,  \pi^2} \,.
\label{erfhcitr}
\eeq
This result is exact, there are no $1/N$ corrections.
Note that all existing calculations of the gluino condensate were done in the
Pauli-Villars (PV) regularization scheme. Equation (\ref{erfhcitr}), as it is,
holds in that scheme. However, as we explain
in Appendix B, for SUSY gluodynamics with dimensional reduction the 
PV scheme  gives the same result as the more currently used  $\overline{\rm MS}$ scheme. 
Thus, all results following from (\ref{erfhcitr}) can be viewed as
referring to the  $\overline{\rm MS}$ scheme.

Now let us turn to the orientifold theory which is planar equivalent to
supersymmetric gluodynamics. The first question to ask
is the mapping of $\left\langle 
\lambda^{a\,,\alpha} \lambda^a_{\alpha}  \right \rangle $
onto $ \langle \bar\Psi_{[ij]} \Psi^{[ij]} \rangle$.
There are many ways to establish a proper normalization.
The simplest way is the comparison of the
corresponding  mass terms. The Dirac fermion $\Psi$  of the
orientifold theory can be replaced by two Weyl spinors, $\xi_{[ij]}$
and $\eta^{[ij]}$, so that the fermion mass term becomes:
\beq
m\bar\Psi\Psi = m\xi\eta + {\rm h.c.}\; ,
\eeq
while in softly broken SYM the mass term has the form
\beq
\frac{m}{2}\, \lambda\lambda + {\rm h.c.}
\eeq
Thus,
\beq
\frac{1}{2} \langle \lambda\lambda\rangle \leftrightarrow 
\langle \xi\eta \rangle\,,\quad\mbox{or}\quad \langle \lambda\lambda\rangle 
\leftrightarrow \langle \bar\Psi\Psi \rangle
\quad\mbox{at}\quad \theta =0\,.
\eeq
The same identification is obtained from comparison
of the two-point functions in the scalar and/or pseudoscalar channels in
both theories.

In the orientifold theory 
\beq
\frac{\gamma}{\beta_0} =\frac{\left(1-\frac{2}{N}\right)\left(1+\frac{1}{N}\right)}{1+\frac{4}{9N}}
=1 +O(1/N)\,,
\label{odin}
\eeq
and
\beq
\frac{3\beta_1}{\beta_0^2} = \frac{1+\frac{19}{9N}-\frac{4}{3N^3}}{\left(1+\frac{4}{9N}\right)^2}=1 +O(1/N)\,.
\label{dva}
\eeq
The nonperturbative planar equivalence  \cite{ASV1} implies
\beq
R_{\rm OR}(N) = R_{\rm SYM}\, \tilde{K}(1/N)\,,\qquad \tilde{K}(1/N) = 1 + O(1/N)\,,
\eeq
where  the $O(1/N)$ terms in $\tilde K$
reflect deviations from the SYM/OR 
equivalence at non-planar level.

 Expressing the result in terms of the conventional fermion bilinear through Eqs.~(\ref{RGicond}) and (\ref{erfhcitr}) we arrive at
\begin{eqnarray}
 \langle \bar\Psi_{[ij]} \Psi^{[ij]} (\mu) \rangle & =& 
-  \frac{ N^2}{2 \pi^2}\,\, \mu^3 \, \Big( \lambda 
  (\mu)\Big)^{- (\gamma/\beta_0 )- 3 (\beta_1/\beta_0^2)}\,\, \exp\left( - \frac{3 N}{\beta_0 \,\,  \lambda (\mu ) } \right) \nonumber\\[3mm]
 &\times&  \tilde{K}(1/N)  \,\, c^{1- (3 \beta_1/\beta_0^2)}\,\,  \tilde{c}^{1- (\gamma/\beta_0)} 
\label{finalexact}
\end{eqnarray}
where all quantities refer to those of $\rm{QCD}_{\rm{OR}}$,
see Eqs (\ref{odin}) and (\ref{dva}).

This is our final general result. It   shows a dependence on the choice of $c$ 
and even more so on $\tilde{c}$ (since
its  exponent is of order $1/2$). Such dependence can be absorbed, however,  in the definition of
the factor $\tilde{K}(1/N)$ modifying just the sub-leading terms.
We, thus, rewrite (\ref{finalexact}) in a simpler form
\begin{eqnarray}
 \langle \bar\Psi_{[ij]} \Psi^{[ij]} (\mu) \rangle &=&
 - \frac{ N^2}{2 \pi^2}\,\, \mu^3 \, \Big( \lambda 
  (\mu)\Big)^{- (\gamma/\beta_0 )- 3 (\beta_1/\beta_0^2)} \exp\left( - \frac{3 N}{\beta_0 \,\,  \lambda (\mu ) } \right) \,\,  {K}(1/N)  \,,\nonumber\\[5mm]
  K(1/N) &=& 1 + O(1/N)\,.
\label{finalredef}
\end{eqnarray}

\section{Finite-$N$ corrections and numerical results}
\label{finite}

\noindent
The fact that $\rm{QCD}_{\rm{OR}}$ goes into YM theory at $N=2$ implies the vanishing of the fermion condensate at $N=2$. In other words we know for sure that the function $K(1/N)$ (as well as the previously introduced $\tilde{K}(1/N)$) must have a zero at $N=2$. Moreover, arguments can be given that this zero is of the first order. Then we can write
\beq
K(1/N) = \left(1-\frac{2}{N}\right) \, K_*(1/N)\,,
\eeq
where $K_*(1/N)$ is supposed to be free from ``large" $1/N$
corrections. 
Assuming that $K_*(1/3)$ differs from 1 by $\pm 30\%$ at most,
and   setting   $N=3$, we arrive at the final formula for the 
quark condensate in one-flavor QCD
\beq
\frac{\langle \bar\Psi_{[ij]} \Psi^{[ij]} (\mu) \rangle}{\mu^3} = 
- \frac{3}{ 2\pi^2} \,\, K_*(1/3)\,   \Big(\lambda (\mu )\Big)^{-1578/961} 
\,\, \exp\left(-\frac{27}{31\, \lambda (\mu )}\right)\,,
\eeq
where $\lambda (\mu ) = 3 \alpha_s(\mu)/2 \pi$, see Eq.~(\ref{thooftc}).

As has been already mentioned, we expect non-planar  corrections
in $K_*$ to be in the ballpark $\pm 1/N$. If so, three values
for $K_*(1/3)$,
\beq
K_*(1/3) = \{ 2/3,\,\, 1,\,\, 4/3\}
\label{set}
\eeq
give a representative set. The only thing we need now
 is the value of $\lambda (\mu )$. Given $\mu$,  
 $\lambda_{\overline{\rm MS}} (\mu )$
and Eq.~(\ref{set}) one can
get a numerical evaluation of the predicted
quark condensate in one-flavor QCD.

The problem is that one-flavor QCD is different   both from real QCD, with three
massless quarks, and from quenched QCD 
in which   lattice measurements have been recently carried out~\cite{lattice}.
In   quenched QCD there are no quark loops in the running
of $\alpha_s$; thus, it runs steeper than in one-flavor QCD.
On the other hand, in three-flavor QCD  the running
of $\alpha_s$ is milder than in one-flavor QCD.

To estimate the input value of  $\lambda_{\overline{\rm MS}} (\mu )$
we resort to the following procedure.
First, starting from $\alpha_s (M_\tau )=0.31$ (which is close to the
world average) we determine  $\Lambda^{(3)}_{\overline{\rm MS}}$.
Then, with this $\Lambda$ used as the input, we evolve  the coupling constant back
to 2 GeV according to the {\em one-flavor} formula. In this way we obtain
\beq
\lambda (2\,\,\rm GeV) = 0.115\,.
\label{perots}
\eeq
Then 
\beq
\langle\bar\Psi\Psi\rangle = -\{0.014,\,\,\, 0.021,\,\,\, 0.028\}\,\,{\rm GeV}^3\,,\qquad
\mu = 2\,\,{\rm GeV}\,,
\label{ournumber}
\eeq
corresponding to three values of $K_*$ in Eq.~(\ref{set}).

A check exhibiting the sensitivity
of our prediction to the value of $\lambda (2\,\,\rm GeV)$
is provided by lattice measurements.
Using the results of Ref.~\cite{Luscher:1993gh} referring to {\em pure} Yang-Mills
theory one can extract $\alpha_s (2\,\,\rm GeV)=0.189$.
(Here and below everything is in $\overline{\rm MS}$.)
Then, as previously, we find  $\Lambda^{(0)}_{\overline{\rm MS}}$,
and evolve back to 2 GeV according to the {\em one-flavor} formula.
The result is
\beq
\lambda (2\,\,\rm GeV) = 0.097\,.
\label{vtorots}
\eeq
The estimate (\ref{vtorots}) is smaller than (\ref{perots})
approximately by one $\sigma$. This is natural since the lattice
determinations of $\alpha_s$ lie on the low side, within one
$\sigma$ of the world average. Using Eq.~(\ref{vtorots})
we would get
then 
\beq
\langle\bar\Psi\Psi\rangle = -\{0.05,\,\,\, 0.07,\,\,\, 0.09\}\,\,{\rm GeV}^3\,,\qquad
\mu = 2\,\,{\rm GeV}\,,
\label{ournumberp}
\eeq

Now we have to compare our prediction with an ``empiric" value 
of the quark condensate in one-flavor QCD.
Chiral perturbation theory allows one to determine
the quark masses (see e.g. \cite{Leutwyler:1996eq}). The Gell-Mann-Oakes-Renner
(GMOR) relation \cite{Gell-Mann:rz} then implies
\beq
\langle \bar\Psi\Psi\rangle = - 0.015\pm 0.005\,\,{\rm GeV}^3\,,
\qquad
\mu = 2\,\,{\rm GeV}\,.
\eeq
One should remember that the very basis of this derivation, the GMOR
relation, implies three light flavors. One can hope, though, that this
particular quantity, $\langle \bar\Psi\Psi\rangle $, is not very sensitive
to the number of light flavors, although it is difficult to assign
any uncertainty associated with the $N_f$ dependence.

Lattice measurements of $\langle \bar\Psi\Psi\rangle $ were performed~\cite{lattice} 
in quenched QCD. Two methods were used.
The first determination was based on the measurement of the strange
quark mass through a fit of the $K$-meson mass to its empiric
value. Then $\langle \bar\Psi\Psi\rangle $ was extracted from the GMOR
relation. The second determination was a direct measurement of
$\langle \bar\Psi\Psi\rangle $. Both methods agree as far as the central
value is concerned, while the uncertainties are much larger in
the latter method. We quote here the result ~\cite{lattice} obtained
in the first method,
\beq
\langle \bar\Psi\Psi\rangle = - 0.019 \pm 0.004\,\,{\rm GeV}^3\,,
\qquad
\mu = 2\,\,{\rm GeV}\,.
\eeq
If, instead, one uses Eq. (47) of Ref.~\cite{lattice}
and substitutes there $a$ in physical units from Ref.~\cite{Luscher:1993gh},
one gets 
\beq
\langle \bar\Psi\Psi\rangle = - 0.012 \pm 0.004\,\,{\rm GeV}^3\,,
\qquad
\mu = 2\,\,{\rm GeV}\,.
\eeq
 In view of the above, it is not unreasonable to assume that
the quark condensate in one-flavor QCD lies between these   values.
Our educated guess is
\beq
\langle \bar\Psi\Psi\rangle_{\rm one-fl\,\, QCD} = - 0.016 \pm 0.005\,\,{\rm GeV}^3\,,
\qquad
\mu = 2\,\,{\rm GeV}\,.
\eeq

 Comparison with Eq.~(\ref{ournumber}) exhibits a significant overlap!
 Given all uncertainties involved in our numerical estimates,
 both from the side of supersymmetry/planar equivalence and the ``empiric" side,
 we can state with satisfaction that the agreement is quite remarkable.
 
 \section{Conclusion}
 
 We started from supersymmetric gluodynamics where powerful methods, such 
 as holomorphy, allow one to exactly calculate the gluino condensate. We
 then  applied the non-perturbative planar equivalence
 obtained in \cite{ASV1} in conjunction with the
 orientifold large-$N$ expansion
 \cite{ASV2} to predict the value of the quark condensate in one-flavor massless QCD,
 up to subleading $1/N$ corrections. This seemingly  first
 {\em quantitative} application of a $1/N$ expansion in $D=4$ produces
 a value for the quark condensate in remarkable agreement with
 the ``empirical value."  Hopefully, this may pave the way to a whole new line of research based on translating a variety of
exact results in supersymmetric theories to ordinary
one-flavor QCD. 

Two extensions of our method look worth being considered:
\begin{itemize} 
\item Evaluation of subleading 
 $1/N$ corrections, for which we were only able to give  rough estimates here;
\item Extension of our method in the direction of connecting non-supersymmetric
or ${\cal N}=1$ supersymmetric theories to
${\cal N} \ge 2$ theories for which even more is known.
\end{itemize}
Whether or not either one of these developments can be carried out  remains to be seen.
 
 \newpage
 
\Acknowledgements

We would like to thank
 Leonardo Giusti for informing us of recent lattice results and to  Valya Khoze for clarifying to us the definition of $\Lambda$ in Ref.~\cite{Davies:1999uw}. 
Special thanks go  to Andre Hoang for bringing
to our attention Ref.~\cite{Hasenfratz:1981tw}, to
Philippe de Forcrand who deciphered
for us Refs.~\cite{lattice,Luscher:1993gh} and to Arkady Vainshtein for 
discussions of   the Pauli-Villars vs. $\overline{\rm MS}$ schemes.
We are grateful to A. Czarnecki, K. Chetyrkin and A. Kataev for communications.

A.A. thanks the
Albert-Einstein-Instit\"{u}t for warm hospitality.
The work  of  M.S. was
supported in part by DOE grant DE-FG02-94ER408.

 \section*{Appendix A. Master formulae}

\renewcommand{\theequation}{A.\arabic{equation}}
\setcounter{equation}{0}
 
In this Appendix we present basic formulae which are repeatedly used as an input
in the bulk of the paper.

The master formula for the gluino condensate in
SU($N$) supersymmetric gluodynamics is
\begin{eqnarray}
\langle  \lambda^{a}_\alpha \lambda^{a\,\alpha}\rangle  &=&
- 32\pi^2 \, M_{\rm PV}^3\, \frac{1}{g^2}\, \exp\left(-\frac{8\pi^2}{Ng^2}
\right)\\[4mm]
 &=& - 4N\,M_{\rm PV}^3\,\frac{1}{\lambda}\, e^{-1/\lambda} \,,
\label{glcond}
\end{eqnarray}
where $M_{\rm PV}$ is the mass of the Pauli-Villars regulator,
$g^2$ is the coupling constant at $M_{\rm PV}$, the gluino field is normalized in such a way that the fermion part of the Lagrangian is ( assuming that the vacuum angle $\theta =0$)
\beq
{\cal L}_{\rm ferm} = \frac{i}{g^2}\,\bar\lambda^{\dot\alpha}\, D_{\dot\alpha\alpha}\lambda^\alpha \,.
\eeq

For numerical comparisons we  need to know $\Lambda^{\rm one-fl}_{\rm QCD}$.
This quantity is estimated in a number of ways in Sect.~\ref{finite}.
We use the standard formala \cite{Hinchliffe} for the running
gauge coupling constant at two loops,
\beq
\alpha (\mu) 
=\frac{4\pi}{\beta_0\, \ln \frac{\mu^2}{\Lambda^2}}\,
\left( 1-\frac{2\beta_1}{\beta_0^2}\, \frac{\ln \ln 
\frac{\mu^2}{\Lambda^2}}{\ln 
\frac{\mu^2}{\Lambda^2}} + ...
\right)\, .
\label{2const}
\eeq
The corresponding expression for $\lambda$ is quoted in Eq.~(\ref{lam}).

 \section*{Appendix B. The Pauli-Villars vs. $\overline{\rm MS}$ regularization schemes}

\renewcommand{\theequation}{B.\arabic{equation}}
\setcounter{equation}{0}

All existing calculations of the gluino condensate are performed in
the Pauli-Villars scheme, while all perturabative calculations
and experimental determinations are routinely carried out
in the  $\overline{\rm MS}$  scheme. Therefore, for our purposes
it is necessary to know the relations between the
corresponding $\alpha$'s or $\Lambda$'s. 

The first derivation of this relation can be found in 't Hooft's
pioneering paper \cite{th}, see Sect. 13. Unfortunately, the key expression
(13.7) contained an error which, unfortunately, propagated in part in some reviews, e.g.
Ref.~\cite{abc}.  It was corrected by Hasenfratz and Hasenfratz
\cite{Hasenfratz:1981tw}, see also Ref.~\cite{Shore:1978eq}, as well as in a 
later reprint of Ref.~\cite{th} (see \cite{msth}). In pure Yang-Mills
theory
\beq
\Lambda_{\rm PV} = \Lambda_{\overline{\rm MS}}\,\exp\left(\frac{1}{22}\right )\,.
\eeq
The difference is entirely due to the fact that the vectorial index $\mu$ of
the gauge connection $A_\mu^a$ takes $D$ rather than four
values in $D$ dimensions. In QCD with $N_f$ flavors
$1/22$ must be replaced by $\left(22-4\frac{N_f}{N}\right)^{-1}$.

In supersymmetric theories the situation is slightly more 
complicated on the one hand, and considerably simpler, on the other.
Indeed, dimensional regularization {\em per se} cannot be used
since it breaks the balance between the number
of the fermionic and bosonic degrees of freedom.
For instance, in SUSY gluodynamics the standard dimensional regularization would  effectively imply
$D-2$ bosonic degrees of freedom and $2$ fermionic.

The problem is fixed by using {\em dimensional reduction}
with the subsequent application of the $\overline{\rm MS}$ procedure.
The supersymmetry is maintained because even in $D\neq 4$
dimensions the numbers of the  fermionic and bosonic degrees of freedom
match.

The most crucial point can be expressed as follows.  In the 't Hooft language
the difference between the PV and $\overline{\rm MS}$ schemes comes entirely
from the non-zero mode parts of the determinants ({\em quantum corrections}
in the instanton background). If one uses a more straightforward
perturbative language, one can split the calculation of the gauge coupling
renormalization (by virtue of
the background field method, in a weak background) in two parts ---
the one associated with the magnetic interaction
with the background field, and the one  associated with the charge interaction.
It is easy to see that the magnetic part produces no difference
between PV and $\overline{\rm MS}$. The difference is entirely
due to the charge part, which is in one-to-one correspondence
with the {\em non-zero mode parts} of the determinants
in the instanton calculation.

For non-supersymmetric theories one must carry out  a special
dedicated calculation
to analyze the difference between  PV and $\overline{\rm MS}$.
In supersymmetric theories the charge-interaction part
in the gauge coupling renormalization cancels (by the same token
all non-zero mode determinants in the instanton background cancel).
This cancellation is due to the balance between the number
of fermionic and bosonic degrees of freedom.
Thus, dimensional reduction plus $\overline{\rm MS}$ procedure
give rise to the same $\Lambda$ as the Pauli-Villars regularization,
\beq
\Lambda_{\rm PV} = \Lambda_{\overline{\rm MS}}\,,\qquad
\mbox{SUSY dimensional reduction}\,.
\eeq

\end{document}